\documentclass[a4paper]{JHEP3}
\usepackage{graphicx}
\usepackage{subfigure}
\usepackage{epsfig}
\newcommand{\be}{\begin{eqnarray}}
\newcommand{\ee}{\end{eqnarray}}

\newcommand{\br}{\bf r}
\newcommand{\bs}{\bf s}
\newcommand*{\Dsl}[0]{{\rlap{\kern2.25pt /}{D}}}

\title{Topological Aspects of Fermions on a Honeycomb Lattice}
\author{{\bf Dipankar Chakrabarti\thanks{Present address: Department of Physics, Indian Institute of Technology Kanpur, Kanpur-208016, India}, Simon Hands and Antonio Rago}
\\ Department of Physics, Swansea University, 
Singleton Park, Swansea, SA2 8PP, UK.}

\date{\today}
\abstract{
We formulate a model of relativistic fermions moving in two Euclidean
dimensions based on a tight-binding model of
graphene. The eigenvalue spectrum of the resulting Dirac operator is solved
numerically in smooth U(1) gauge field backgrounds carrying an integer-valued
topological charge $Q$, and it is demonstrated that the resulting number of
zero-eigenvalue modes is in accord with
the Atiyah-Singer index theorem applied to two continuum flavors. 
A bilinear but gauge non-invariant
chirality operator appropriate for
distinguishing the topological zero modes is identified. 
When this operator is used to calculate $Q$, it is found that
the maximum topological charge capable of being measured in this fashion
scales with the perimeter of the lattice. Some concluding remarks compare 
these results
to what is known for staggered lattice fermions.
}

\keywords{Lattice gauge field theories, Field theories in lower dimensions}
\preprint{}

\begin{document}

\section{Introduction}
Following pioneering work in the 1980's~\cite{KarstenWilczek}, there has been a
recent revival of interest in lattice fermion formulations with a minimal flavor
content of chirally-symmetric fermions \cite{creutz,borici,bedaque,Cichy}.
Stimulated by the excitations known to obey a 
quasi-relativistic Dirac equation in graphene (a 
monolayer of carbon atoms arranged in a honeycomb lattice which has recently
been realised experimentally), Creutz~\cite{creutz} devised a four-dimensional
Euclidean lattice action describing two species of massless chirally-invariant
fermion, each centred at a special location $\pm\tilde p_\mu$ 
in momentum space.
Creutz's action is formulated on a hypercubic lattice, with tunable
parameters enabling the magnitude of $\tilde p_\mu$ to be controlled.
Bori\c ci~\cite{borici} soon developed the idea, finding an action with two
flavors
located this time at the origin 
and at $({\pi\over2a},{\pi\over2a},{\pi\over2a},{\pi\over2a})$. In each case the
flavor content is the minimum consistent with the celebrated no-go theorem 
governing lattice fermion actions which are local, unitary, and chirally
symmetric~\cite{NN}. This has led to hopes that these formulations could form
the basis for an inexpensive alternative to overlap fermions in realistic
lattice QCD simulations with two light quark flavors.

The feature of these actions is that in the vicinity of the special ``Dirac
points'' the eigenvalues $E$ of a suitably-defined Hamiltonian operator
can be written as $E(p_\mu-\tilde p_\mu)=\pm K\vert z(p)\vert$, 
where $z(p)$ can be written as $z_0
e^{i\theta}$
in two Euclidean dimensions and as $z_0+i\vec z.\vec\sigma$ in four
dimensions~\cite{creutz}.
In 2$d$ $z$ maps an S$^1$ surrounding the Dirac point to a complex phase: in
4$d$ the analogous mapping is from S$^3$ to a quaternionic space. In either case
the solution of the fermion Hamiltonian engenders a non-trivial wrapping,
implying that the surface must enclose a zero, so that
the only consistent result on shrinking it to a
point is $E=0$. Hence the existence of the Dirac points, and the linear nature
of the dispersion $E(p_\mu-\tilde p_\mu)$ in their immediate vicinity, 
is topologically stable.
\footnote{
Because of the periodicity of the Brillouin Zone, the Dirac points must appear
in pairs so that the overall wrapping vanishes~\cite{NN}.}
One might therefore hope that the desirable properties listed above persist even
once interactions with gauge degrees of freedom are introduced.

A difficulty has been pointed out in \cite{bedaque}; the
actions of \cite{creutz,borici}, while chirally symmetric, break 
important hypercubic and 
discrete symmetries such as parity
and time-reversal, meaning that they are not protected against acquiring
non-covariant 
counterterms such as eg. $\bar\psi\gamma_\mu\psi$ through quantum
corrections. This introduces a severe fine-tuning problem to any practical
simulation
programme based on the original actions. Recently Creutz has proposed a
refinement of the action of \cite{borici} in which it is claimed such effects
can be mitigated to the point where perhaps they are manageable~\cite{creutz2}.

The purpose of the current paper is to investigate the interaction of these
minimal chiral fermions with gauge fields, not via a perturbative approach to
radiative corrections~\cite{bedaque,Cichy}, 
but rather in a non-perturbative manner via their
response to a globally-defined topological charge. It is well-known that in a
gauge background  with integer-valued topological charge $Q$, the spectrum of
the Dirac operator $D$ obeys the Atiyah-Singer index theorem
\be 
Q=
n_+-n_-, 
\label{eq:AS1}
\ee 
where $n_{+(-)}$ denotes the number of positive (negative) chirality zero modes
of $D$. We have been inspired by the classic paper \cite{smit_vink} of Smit and
Vink, who studied the extent to which (\ref{eq:AS1}) is obeyed for both
staggered and Wilson lattice fermions. Accordingly, we will investigate the
response of minimal chiral fermions in $2d$ by calculating the spectrum of $D$
in a background U(1) gauge field corresponding to a quantised  homogeneous flux,
which can be shown to carry $Q\not=0$.  Because of the combination of apparent
simplicity with high symmetry, we have chosen to track closely the original
connection with graphene and hence formulate our fermions on a honeycomb
lattice. In $4d$ the analogous ``hyperdiamond'' lattice corresponds to a
particular choice of parameters in \cite{creutz}, and has been explored further
in \cite{bedaque2}. Despite the aesthetic appeal of the honeycomb, it presents
technical challenges which we feel are worth reporting in some detail.
We will show that the index theorem (\ref{eq:AS1}) is indeed satisfied, and 
find a definition of chirality $\bar\psi\sigma_3\psi$ capable of distinguishing
between zero and non-zero modes. The drawback is that it is not gauge-invariant,
and may not even be universal.

The special properties of fermions hopping on a honeycomb lattice have been
known in the literature for many years, ever since 
Semenoff~\cite{Semenoff:1984dq} noted the relativistic nature of
the dispersion around the Dirac points and solved the resulting Dirac equation 
in a homogeneous magnetic field in the long-wavelength limit to expose
topologically-stable zero modes, confirming a result found in continuum field
theory by Jackiw~\cite{Jackiw:1984ji}. More recently, the topological aspects of
honeycomb fermions interacting with an external magnetic field have been used to account for
the unconventional quantum Hall effect in graphene~\cite{QHE}. The index theorem
has also been used to analyse the effects of point defects in
graphene~\cite{Stone}. In these papers the authors either worked in the
continuum limit of the lattice model or worked in momentum space. In this work,
by contrast, we will examine the applicability of the continuum results to
fermions defined on the finite lattices relevant for QCD simulations;
our concerns will be the approach to and recovery of the continuum limit
predictions, and the technical issues associated with defining a chirality
operator referrred to in the previous paragraph.

The remainder of the paper is organised as follows. In
Sec.~\ref{sec:formulation} we specify the honeycomb lattice, define the 
lattice Dirac operator, and show that in the long-wavelength limit an action
describing two continuum Dirac flavors is recovered. Sec.~\ref{sec:index}
reviews the index theorem, and outlines how U(1)
configurations with $Q\not=0$ may 
be constructed on the honeycomb, and Sec.~\ref{sec:anal_spectrum} presents
the Dirac spectrum calculated both for free fermions on the honeycomb, and for
continuum fermions on backgrounds with $Q\not=0$. In Sec.~\ref{sec:results}
we then present numerical results for the spectrum calculated on lattices up to
size $100\times100$ with $Q\not=0$. The definition used for 
the chirality of a mode 
differs from the naive expectation based on free fermions.
Nonetheless, we will show that both the spectrum and the index
calculated on the basis of this chirality match analytic expectations
provided $Q$ is not too large; interestingly, the maximum value of $Q$ for which
continuum results are reproduced turns out to scale with the perimeter of the
lattice. Our concluding remarks in Sec.~\ref{sec:conclusions} will contrast what
we have found with what is known for staggered lattice fermions. Some technical
details concerning the definition of the Fourier transform on a finite honeycomb
lattice are postponed to an Appendix.

\section{Lattice action}
\label{sec:formulation}
\FIGURE[ht]{
\includegraphics[width=8.5cm]{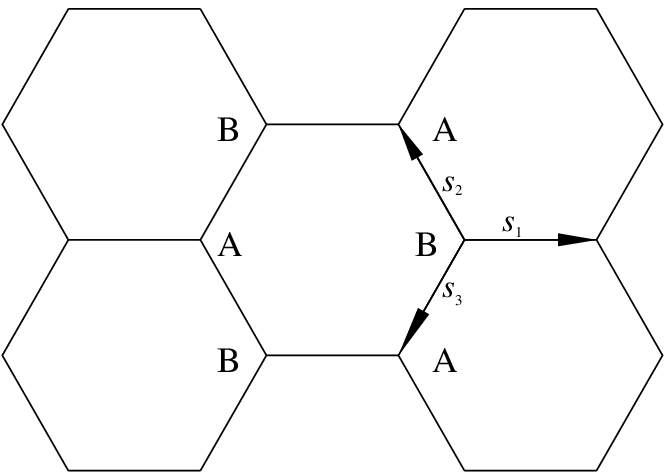}
\label{fig1}
\caption{Honeycomb lattice}
}
It is convenient to begin our presentation using a Hamiltonian devised for
physical (ie. $2+1$-dimensional) graphene~\cite{jackiw_pi}. It is assumed that
on each site {\bf r} 
of a honeycomb lattice there is a mobile electron which may hop to
a neighbouring site under the constraints of the Pauli Exclusion Principle.
Electron spin may be ignored for now;
the tight-binding Hamiltonian is then
\begin{equation}
H=-t\sum_{{\br}\in B}\sum_{i=1}^3
b^\dagger({\br})U({\br},{\bs}_i)a({\br}+{\bs}_i)+
a^\dagger({\br}+{\bs}_i)U^\dagger({\br},{\bs}_i)b({\br}),
\label{eq:H}
\end{equation}
where $t$ is a hopping parameter,
${\bs}_i(i=1,2,3)$ are the three vectors along the 
links as shown in
Fig. \ref{fig1}:
\be
{\bs}_1=(1,0)l,~~~{\bs}_2=\big(-{1\over 2},{\sqrt{3}\over 2}\big)l,
~~{\bs}_3=\big(-{1\over 2},-{\sqrt{3}\over 2}\big)l,
\ee 
and $l$ is the honeycomb bond length. The sites
labelled A and B belong to inequivalent sublattices,
on which 
the operators $a^\dagger\,(a)$ and $b^\dagger\,(b)$ respectively
create (destroy) fermions. The variable 
$U({\br},{\bs}_i)$ is a U(1)-valued gauge connection emerging from the
B site at ${\br}$ along ${\bs}_i$. The Hamiltonian (\ref{eq:H}) is thus
invariant under U(1) gauge transformations. 

To expose the relativistic nature of the low-energy excitation spectrum,
define $H_0=H[U=1]$ and transform to momentum space (we will
refine our definition of the Fourier transformation during the course of what
follows):
\be
H_0=\sum_{\vec k}\left(\Phi(\vec k)a^\dagger(\vec k)b(\vec k)+
\Phi^*(\vec k)b^\dagger(\vec k)a(\vec k)\right) 
\ee
with
\be
\Phi(\vec
k)=-t\left[e^{ik_xl}+2\cos\Bigl({{\sqrt{3}k_yl}\over2}\Bigr)e^{-i{{k_xl}\over2}}\right].
\ee
Consider a basis of Fock states $\vert
\vec k_\pm\rangle=(\sqrt{2})^{-1}[a^\dagger(\vec k)\pm b^\dagger(\vec
k)]\vert0\rangle$ where $a\vert0\rangle=b\vert0\rangle=0$, $\{a^\dagger(\vec
k),a(\vec k^\prime)\}=\delta^2(\vec k-\vec k^\prime)$, 
$\{a,a\}=\{a^\dagger,a^\dagger\}=0$ etc. It is
straightforward to see that $\langle\vec k_\pm\vert H_0\vert\vec k_\pm\rangle
=\pm(\Phi(\vec k)+\Phi^*(\vec k))\equiv\pm E(\vec k)$, and hence that the spectrum is
symmetric about zero. At half-filling (one electron per site) the Fermi energy
is thus at $E=0$.

The dispersion relation $E(\vec k)$ vanishes not at $\vec k=\vec0$, but at the
six corners of the first Brillouin zone, which is also a hexagon but rotated 
by 90$^\circ$ with respect to the cells of Fig.~\ref{fig1}. The corners of this 
hexagon lie at the centres of inequivalent equilateral triangles formed from
reciprocal lattice points;
the Brillouin zone thus
contains two inequivalent {\it Dirac points\/} around which an effective 
low-energy description can be built, which we will take as 
$\vec K_\pm=(0,\pm{{4\pi}\over{3\surd3l}})$. Around these points we can expand:
\be
\Phi(\vec K_\pm+\vec p)=\pm v_F[p_y\mp ip_x]+O(p^2)
\ee
where the Fermi velocity $v_F={3\over2}tl$.
It is now possible, by defining field operators in the neighbourhood of the
Dirac points via $a_\pm(\vec p)=a(\vec K_\pm+\vec p)$ etc, to recast the
Hamiltonian in relativistic form:
\be
H_0\simeq v_F\sum_{\vec p}\Psi^\dagger(\vec p)\vec\alpha.\vec p\Psi(\vec p)
\ee
where $\Psi$ is the column vector $(b_+,a_+,a_-,b_-)^T$ and the $4\times4$ matices
$\vec\alpha$ are defined by
\be
\alpha_x=\left(\matrix{-\sigma_2&\cr&\sigma_2\cr}\right);\;\;\;
\alpha_y=\left(\matrix{\sigma_1&\cr&-\sigma_1\cr}\right)
\ee
so that $\{\alpha_i,\alpha_j\}=2\delta_{ij}$. In this form $H_0$ is easily seen
to be proportional to the Dirac Hamiltonian describing a single massless
four-component
spinor moving with speed $v_F$. 
For physical graphene, the Hamiltonian (\ref{eq:H}) must be modified to
incorporate electron spin; this results in a relativistic $d=2+1$ model with
two four-component flavors.

Starting from the
Hamiltonian (\ref{eq:H}) 
an action for chiral gauge theory in $d=2+1$ was 
proposed by Jackiw and Pi in  \cite{jackiw_pi}.
In this paper
we instead recast it as a $d=2$ Euclidean quantum field theory with 
action of the form $S=\bar\chi D\chi$,
describing two species of fermion field sitting on a
honeycomb lattice, each species occupying a distinct sublattice.
The resulting equation of motion resembles the Dirac equation
in the long wavelength limit; 
the Dirac operator $D$ can be written as 
\be
(D\chi)(x)&=&D_1(x+\hat{0})\chi(x+\hat{0})+D_1(x-\hat{0})\chi(x-\hat{0})
+D_2(x+\hat{2})\chi(x+\hat{2}) \nonumber \\
&&+D_2(x-\hat{2})\chi(x-\hat{2})+ D_3(x)\chi(x).
\label{eq:D}
\ee
In writing the operator this way we have introduced the notion of a {\it
lattice\/} of identical A sites, with rhombus-shaped unit cells of
side $a=\sqrt{3}l$ each
containing one A site and one B site (see Fig.~\ref{fig2}). Each cell is indexed
by a vector $x$, whose form will be specified below, and primitive 
vectors $\hat0={\bs}_1-{\bs}_3$ and $\hat1={\bs}_2-{\bs}_1$ define the lattice
axes. It is also convenient to define the dependent vector $\hat2=\hat0+\hat1$.
The different elements of the operator are then written
\be
D_1(x+\hat{0})&=&\pmatrix{ 0 & 0 \cr U(x,{\bs}_1) & 0}, ~~
D_1(x-\hat{0})=\pmatrix{ 0 & U^*(x-\hat{0},{\bs}_1) \cr 0 & 0},\nonumber\\
D_2(x+\hat{2})&=&\pmatrix{ 0 & 0 \cr U(x,{\bs}_2) & 0}, ~~
D_2(x-\hat{2})=\pmatrix{ 0 & U^*(x-\hat{2},{\bs}_2) \cr 0 & 0},\nonumber\\
D_3(x)&=&\pmatrix{ 0 &  U^*(x,{\bs}_3)\cr U(x,{\bs}_3) & 0},\label{eq:DD}
\ee
with the spinors
\be
\chi(x)=\pmatrix{\chi_A(x) \cr \chi_B(x)}
\ee
where $\chi_A(x)$ and $\chi_B(x)$ are single-component Grassmann fields located
at sites A and 
B of cell $x$ respectively. 

\FIGURE[hb]{
\includegraphics[width=8.5cm]{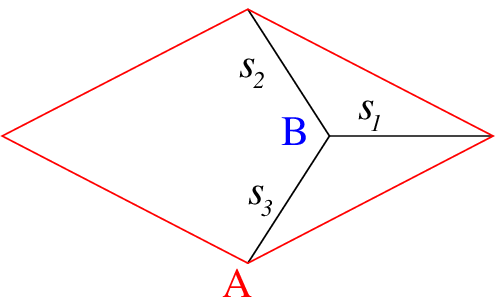}
\caption{The primitive cell}
\label{fig2}
}
Note that with our definition of the link directions all terms in the action are
either of the form $\bar\chi_BU\chi_A$ or $\bar\chi_AU^*\chi_B$.
While the antihermiticity of the Dirac operator (\ref{eq:D},\ref{eq:DD}) is not 
manifest, a little care and close inspection of Figs.~\ref{fig1},\ref{fig2} 
will 
convince the reader that
this is indeed the case. The above analysis expanding $S$ about the Dirac points
$\vec k=\vec K_\pm$ 
goes through as before; writing $\psi_1=(\chi_{B+},\chi_{A+})^T$,
$\bar\psi_1=(\bar\chi_{B+},\bar\chi_{A+})$, 
$\psi_2=(-\chi_{A-},-\chi_{B-})^T$ 
and $\bar\psi_2=(\bar\chi_{A-},\bar\chi_{B-})$, with $\chi_{A\pm}(\vec p)=
\chi_A(\vec K_\pm+\vec p)$ as before,
we obtain
\be
S_0\simeq{3l\over2}\sum_{\vec p}\sum_{\alpha=1}^2\bar\psi_\alpha\vec
p.\vec\sigma\psi_\alpha,
\label{eq:contaction}
\ee
that is, a relativistically covariant action describing two flavors of
two-component spinor moving in $d=2$ Euclidean dimensions,
each flavor localised at one single Dirac point.
Chirality is then {\it naively} defined by the bilinear 
\be
\bar\psi_\alpha\sigma_3\psi_\alpha=\bar\chi_{B+}\chi_{B+}-\bar\chi_{A+}\chi_{A+}-\bar\chi_{A-}\chi_{A-}+\bar\chi_{B-}\chi_{B-}=
\bar\chi_B\chi_B-\bar\chi_A\chi_A,
\label{eq:naivechiral}
\ee
where the second equality assumes that all parts of momentum space can be 
treated uniformly, leading in effect to 
a staggered order parameter.
As we shall see in Sec.~\ref{sec:results} below, 
the definition needs to be modified in the presence of gauge
fields.

\DOUBLEFIGURE[ht]{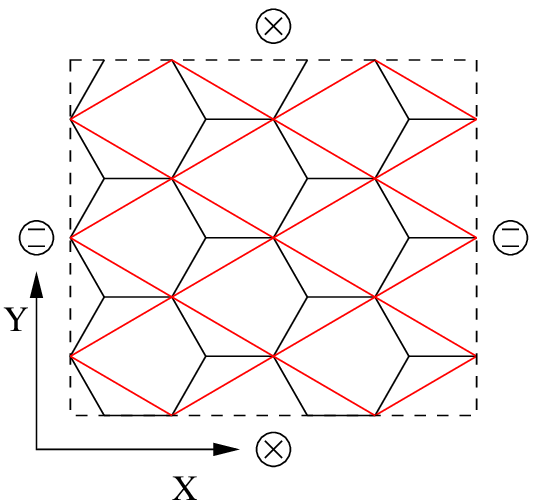,width=5cm}{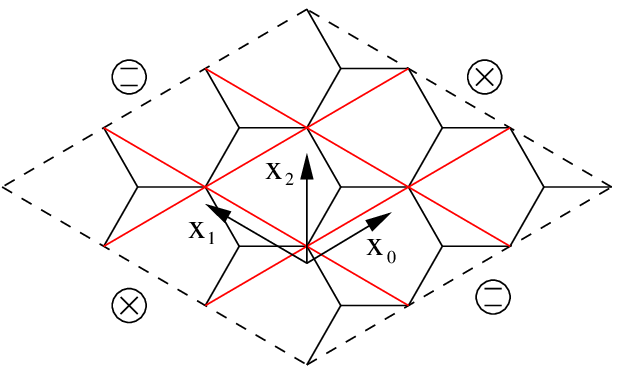,width=5cm}{Perpendicular compactification\label{fig3}}{Primitive compactification\label{fig4}}

To define a finite, translationally-invariant lattice we need to close
the manifold by specifying
boundary conditions. In this paper we have studied two distinct possibilities.
The technically simpler choice is to close the manifold along the two 
non-orthogonal axes $\hat0$ and $\hat1$, for instance defining an
$L_0\times L_1$ system by requiring
$f(x+L_0\hat0)=f(x+L_1\hat1)=f(x)$. We call this the ``primitive'' lattice shown
in red in Fig.~\ref{fig4}. Note it contains $L_0L_1$ distinct hexagons.
We 
can alternatively choose to implement 
the boundary conditions along the orthogonal axes $\hat X$ and $\hat Y$,
in this case calling it 
the ``perpendicular'' lattice shown in black in
 Fig. \ref{fig3}. Some care is needed in indexing the lattice this way: 
it is
convenient to assign the two A sites ${\br}$ and ${\br}-{\bs}_3+{\bs}_2$ and 
the B site ${\br}-{\bs}_3$ the same $\hat X$ index, but to assign them $\hat Y$
indices of respectively eg. 0, $1\over2$, and 1. In this way a lattice which
extends $L_X$ units along $\hat X$ and $L_Y$ units along $\hat Y$ contains
$L_XL_Y$ distinct hexagons.

\section{Index theorem}
\label{sec:index}
Let us consider fermions in the presence of a topological 
charge $Q$ created by 
a background gauge field configuration.
In the continuum, the Atiyah-Singer index theorem relates the topological charge
to the number of chiral zero modes of the fermion. If $\psi_i$ are the
eigenstates of the antihermitian Dirac operator with eigenvalue $iE_i$,
then 
\be Q=\sum_{i,E_i=0}\psi_i^\dagger
\gamma_5\psi_i=n_+-n_-, 
\label{eq:AS}
\ee 
where $n_{+(-)}$ is the number of zero eigenvalue modes of
positive (negative) chirality, ie. satisfying $\gamma_5\psi_i=\pm\psi_i$.  
A heuristic derivation of this relation for
lattice fermions is given in \cite{smit_vink}.  Since chiral symmetry is
minimally broken, we anticipate the above relation holds good on a honeycomb
lattice with a suitable generalisation of the chirality operator $\gamma_5$. 

In two Euclidean dimensions topological charge density is proportional to
the magnetic field strength tensor $F_{12}$. The two-dimensional analogues of 
instantons are localised vortices carrying a quantised magnetic flux; 
the topological charge
$Q$ is defined by 
\be
Q={1\over 2 \pi}\int d^2x F_{12}.
\label{eq:quant}
\ee
On a finite system it is also possible to define homogeneous backgrounds 
with $Q\not=0$.
Consider the 
abelian field strength tensor $F_{12}=\partial_x A_y-\partial_y A_x$ on a 
$2d$ system with boundaries closed in orthogonal directions.
Following \cite{smit_vink}, we then choose  
$A_x(x,y)=-\omega y$ and $A_y(x,y)=0$, so that $F_{12}=\omega$. On a 
$L_x\times L_y$ square lattice of spacing $a$ the 
gauge field $A_1$ at the boundary $y=L_ya$ is related to that at $y=0$ by a 
gauge transformation \cite{smit_vink} 
\be
A_x(y=0)=A_x(y=L_ya)+i\Omega_y\partial_x\Omega_y^{-1}
\ee
where 
\be
\Omega_y(x,y)=e^{i \omega L_y ax}.
\label{eq:transition}
\ee
The discontinuity is permitted
since 
any gauge invariant object remains 
continuous across the boundary of the lattice. Demanding periodicity in the
$x$-direction as well
results in a field strength quantised consistently with (\ref{eq:quant}):
\be
\omega={{2\pi}\over{L_xL_ya^2}}Q.
\label{eq:Qquant}
\ee

On the honeycomb lattice 
the gauge background with constant field strength $\omega$ over 
the lattice, and accordingly equal flux $\omega{\cal A}$ through each hexagonal 
plaquette of area ${\cal A}={{\surd3}\over2}a^2$, 
is quantised according 
to
\be
\omega={4 \pi\over\surd{3} L_x L_ya^2}Q.
\ee
The link field configuration depends 
on which boundary condition we consider. For the 
primitive lattice boundary condition, a possible choice is: 
\be
U(x,{\bs}_1)=\exp\left(-i{{\surd3}\over2}\omega x_{0}a^2\right);
~~U(x,{\bs}_2)=1;
~~U(x,{\bs}_3)=1,
\ee
for all cells except those of the last row with $x_{0}=L_0-1$ 
where in addition
we require
\be
U(x_{0}=L_0-1, x_{1},{\bs}_3)
=\exp\left(-i{{\surd3}\over2}\omega L_0 x_{1}a^2\right).
\ee
It is readily checked that each hexagonal plaquette then has the value
$\exp(i\omega{\cal A})$. 

For the perpendicular boundary condition, 
a link field configuration for the same constant field strength $\omega$ 
could be chosen as follows.
For $x_{Y}=0,{1\over2}, 1,\ldots,L_Y-{1\over2}$:
\be
U(x,{\bs}_1)=\exp\left(-i{{\surd3}\over2}\omega x_{Y}a^2\right);~~
~~U(x,{\bs}_2)=1;
~~U(x,{\bs}_3)=1,
\ee
whereas for the links in the last row we need
\be
U(x_{Y}=L_Y-{\textstyle{1\over2}},x_{X},{\bs}_2)
=\exp\left(i{{\surd3}\over4}\omega L_Y x_{X}a^2\right);\nonumber\\
U(x_{Y}=L_Y,x_{X},{\bs}_3)
=\exp\left(i{{\surd3}\over4}\omega L_y x_{X}a^2\right).
\ee

\section{Dirac spectrum: analytic results }
\label{sec:anal_spectrum}
With the above gauge field configuration we need to solve the lattice
Dirac equation
\be
D \chi_i=iE_i \chi_i
\ee
with  $\chi_i(x)$ satisfying primitive or perpendicular boundary conditions as 
appropriate.
For an arbitrary gauge field background we can do only that numerically. 
In this section we first discuss the 
spectrum on a honeycomb lattice without any gauge field, 
and then for a continuum Dirac operator on a 
smooth background field with $Q\not=0$ of the kind outlined in
Sec.~\ref{sec:index}. 

If we consider the boundary condition on the primitive lattice and
in the free field limit, a plane wave ansatz gives the eigenvalues 
\be
E(k_{0},k_{1})=
\pm \sqrt{3+2 \cos k_{0}+2\cos(k_{0}-k_{1})+2\cos k_{1}}.
\label{EV}
\ee
For  periodic boundary conditions along $\hat0$ and $\hat1$, 
the allowed 
momentum modes are given by
\be
k_{0}&=&{2 \pi n_0\over L_0},~~n_0=0,1,2, \cdots L_0-1;\nonumber \\
k_{1}&=&{2 \pi n_1\over L_1},~~n_1=0,1,2, \cdots L_1-1.
\ee
Note that $k_{0}\hat0$ and  $k_{1}\hat1$ are not orthogonal. 
One can clearly see that 
$E(k_{0},k_{1})$ is not minimised at $k_{0}=k_{1}=0$,
but rather at the Dirac points $K_\pm=(\mp{2\pi\over 3},
\pm{2\pi\over3})$: 
recovering the 
continuum relativistic dispersion relation is therefore nontrivial. 
Writing $k=K_\pm+q$, 
the leading term in the expansion is
\be
E(q_{0},q_{1})=\pm \sqrt{q_{0}^2-q_{0}q_{1}+q_{1}^2}.
\ee
Replacing the non-orthogonal $q_{0},~q_{1}$ by orthogonal 
momenta $p_x,~p_y$ via $q_{0,1}=p_x\pm{1\over\surd3}p_y$, 
we get the desired
relativistic dispersion relation
\be 
E(p_x,p_y)=
\pm \sqrt{{p_x}^2+{p_y}^2}.
\label{eq:cont_disp}
\ee
If expressions (\ref{EV}-\ref{eq:cont_disp}) are 
required in terms of dimensionful momenta $\tilde k, \tilde q, \tilde p$, then
note that $p=a\tilde p$ etc.

The solution for a single flavor of Dirac fermion moving in a $2d$
uniform background magnetic
flux density $\omega$ was first discussed in \cite{Jackiw:1984ji}; here we 
review the explict solution 
given in \cite{smit_vink}. The equation is

written
\be D\psi_j=\sum_{\mu=1}^2D_\mu\sigma_\mu\psi_j=iE_j\psi_j,
\ee
where $\psi$ is a two-component spinor. For a background with topological charge
$Q\not=0$ 
there are $\vert Q\vert$ independent solutions of the form (with $a=1$):
\be
\psi_{n\pm}(x,y)\propto\!\!\!\sum_{\ell=-\infty}^{\infty}e^{2\pi i{x\over
L_x}(j+\ell\vert Q\vert)}e^{-{1\over2}\vert\omega\vert(y\pm{{L_y}\over
\vert Q\vert}(j+\ell\vert Q\vert))^2}\!
H_n\!\!\left(\sqrt{\vert\omega\vert}(y\pm{{L_y}\over{\vert Q\vert}}(j+\ell\vert
Q\vert))\right)\phi_\pm,
\ee
where $j=0,1,\ldots,\vert Q\vert-1$, $\phi_+=
\scriptstyle{\left(\matrix{0\cr1\cr}\right)}$,
$\phi_-=\scriptstyle{\left(\matrix{1\cr0\cr}\right)}$, $H_n$ are Hermite
polynomials of order $n$,
and $Q$ and $\omega$ are related via
(\ref{eq:quant}).  The corresponding eigenvalues are given by
\be 
E^2_{n\pm}=(2n+1)\vert\omega\vert\mp\omega.
\ee
Rearranging, we find a spectrum 
\be
E^2_m=2m\vert \omega\vert,\;\;\;m=0,1,2,\ldots
\label{eq:spec_anal}
\ee
with degeneracy
\be
g_m=\cases{\vert Q\vert&$m=0$;\cr2\vert Q\vert&$m>0$.}
\label{eq:degen_anal}
\ee
The $\vert Q\vert$ zero modes are all proportional to $\phi_+$ ($\phi_-$) for
$Q$ positive (negative), in accordance with the index theorem (\ref{eq:AS}).
For $m>0$ an equal number of positive and negative chirality solutions can be 
found. The increase of $g_m$ with $\omega$ is a
relativistic analogue of the Landau levels observed in metals in a strong
magnetic field. For the two continuum flavors described by the honeycomb Dirac
operator (\ref{eq:D}), 
the index theorem thus predicts $2\vert Q\vert$ zero modes, a result
first obtained in \cite{Semenoff:1984dq}.

\section{Numerical Results}
\label{sec:results}
In order to analyze the spectrum of the Dirac operator in various gauge field
backgrounds, the matrix $-D^2[U]$ 
was diagonalised 
via a subspace iteration technique,
using  Chebyshev polynomial iteration to accelerate the 
convergence of the eigenvalues $E^2$. 
Since small eigenvalues converge at a faster rate
than the high lying eigenvalues, locking the already converged
eigenvalues and eigenvectors also accelerates the convergence 
of the other eigenvalues. The locked eigenspaces are only used to 
orthogonalise the remaining subspaces. This algorithm is also suitable to 
find the few lowest lying eigenvectors.
Further details may be found in 
\cite{saad, luscher}. 

\FIGURE[ht]{
\includegraphics[width=12.0cm]{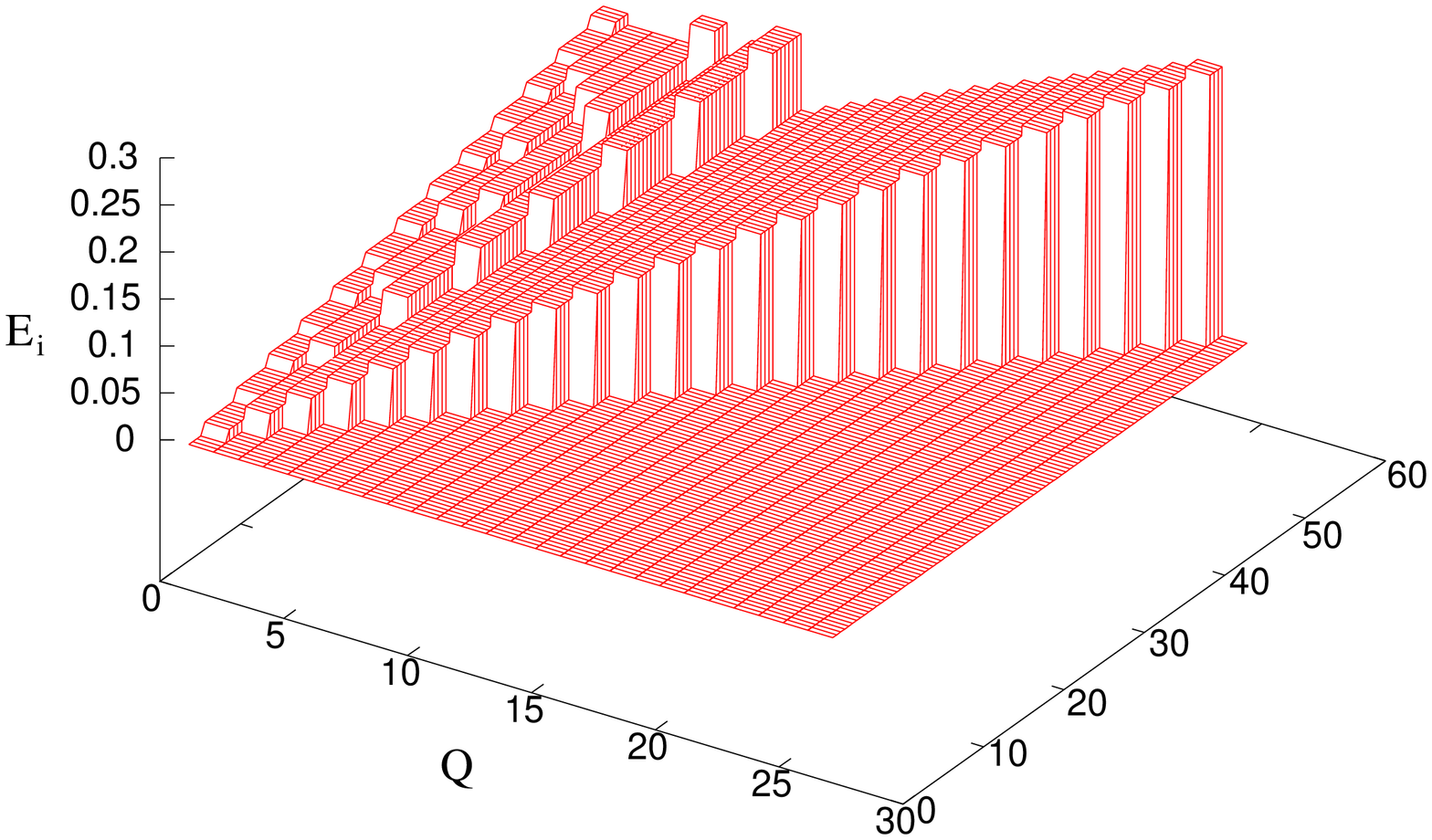}
\caption{Eigenvalue spectrum as a function of $Q$ on a $30\times30$ lattice
(the vertical scale is measured in units where $a^{-2}={{\surd3}\over2}$).}
\label{fig:spectrum}
}

In Fig.~\ref{fig:spectrum} 
we plot the 60 smallest eigenvalues $E^2$ calculated on
a $30\times30$ lattice with primitive boundary conditions, for values of
topological charge $Q$ varying between 1 and 26. The spectrum for the
perpendicular
boundary conditions is identical. 
On this lattice $\omega
a^2=0.00698Q$. Close inspection of the figure reveals very good agreement with 
both the eigenvalue 
prediction (\ref{eq:spec_anal}) and the degeneracy pattern
(\ref{eq:degen_anal}), recalling that for two flavors we expect a degeneracy 
$2g_m$.
In particular, the triangular ``carpet'' with
$E^2=0$ corresponds to the zero modes with degeneracy growing linearly with $Q$
as predicted by the index theorem (\ref{eq:AS}). It is important to note that
for these smooth background configurations the zero-mode eigenvalue is equal to
zero within machine precision, just as is the case for staggered
fermions~\cite{smit_vink}. In what follows we will
strengthen this correspondence by specifying 
a chirality operator appropriate for
honeycomb fermions.

\FIGURE[ht]{
\centering
\mbox{\subfigure[$E=0$, Multi-valued FT, A sublattice]
{\includegraphics[width=5.0cm,angle=-90]{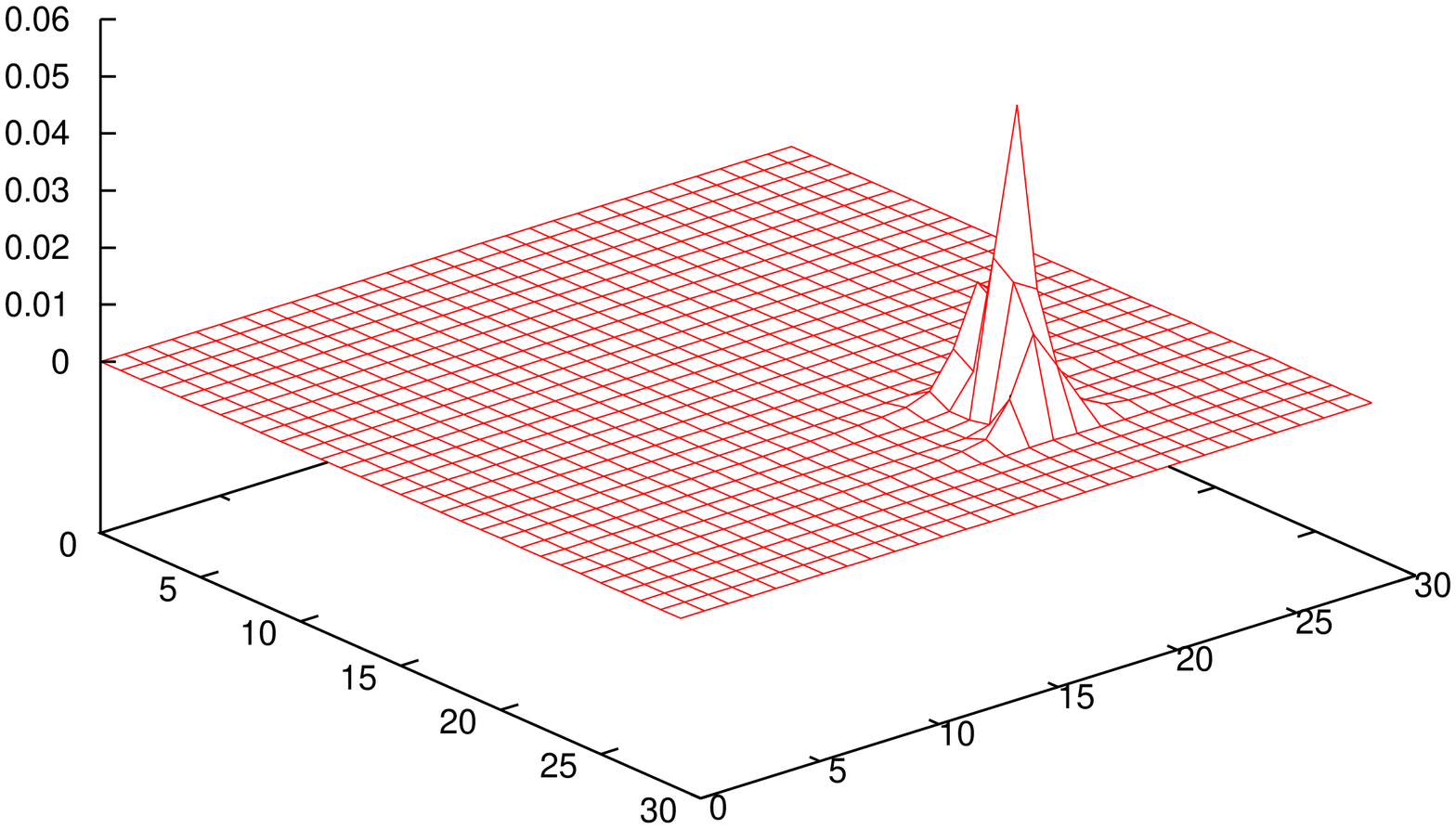}}
\quad 
\subfigure[$E=0$, Multi-valued FT, B sublattice]
{\includegraphics[width=5.0cm,angle=-90]{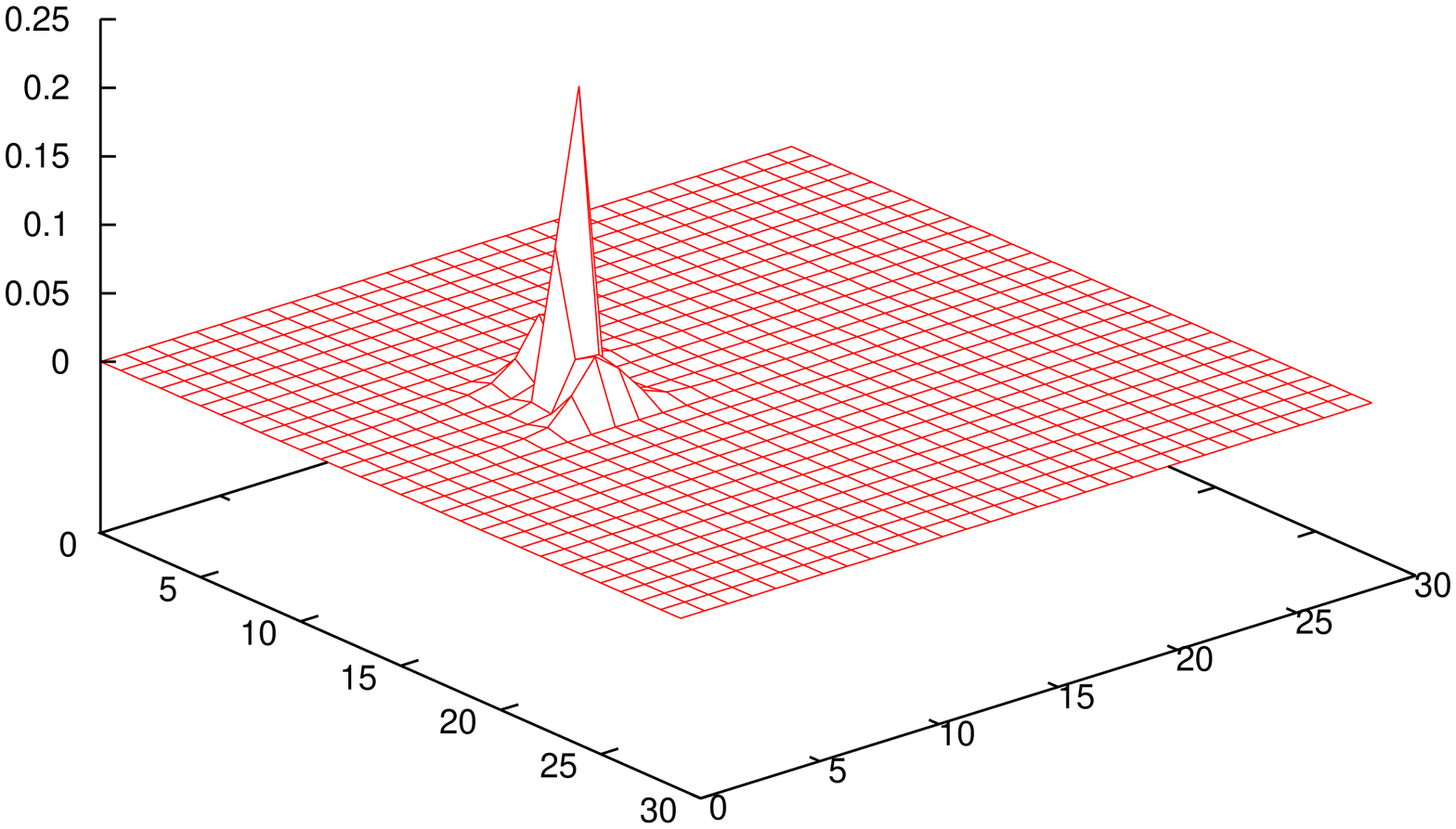}}}
\mbox{
\subfigure[$E\not=0$, Multi-valued FT, A sublattice]
{\includegraphics[width=5.0cm,angle=-90]{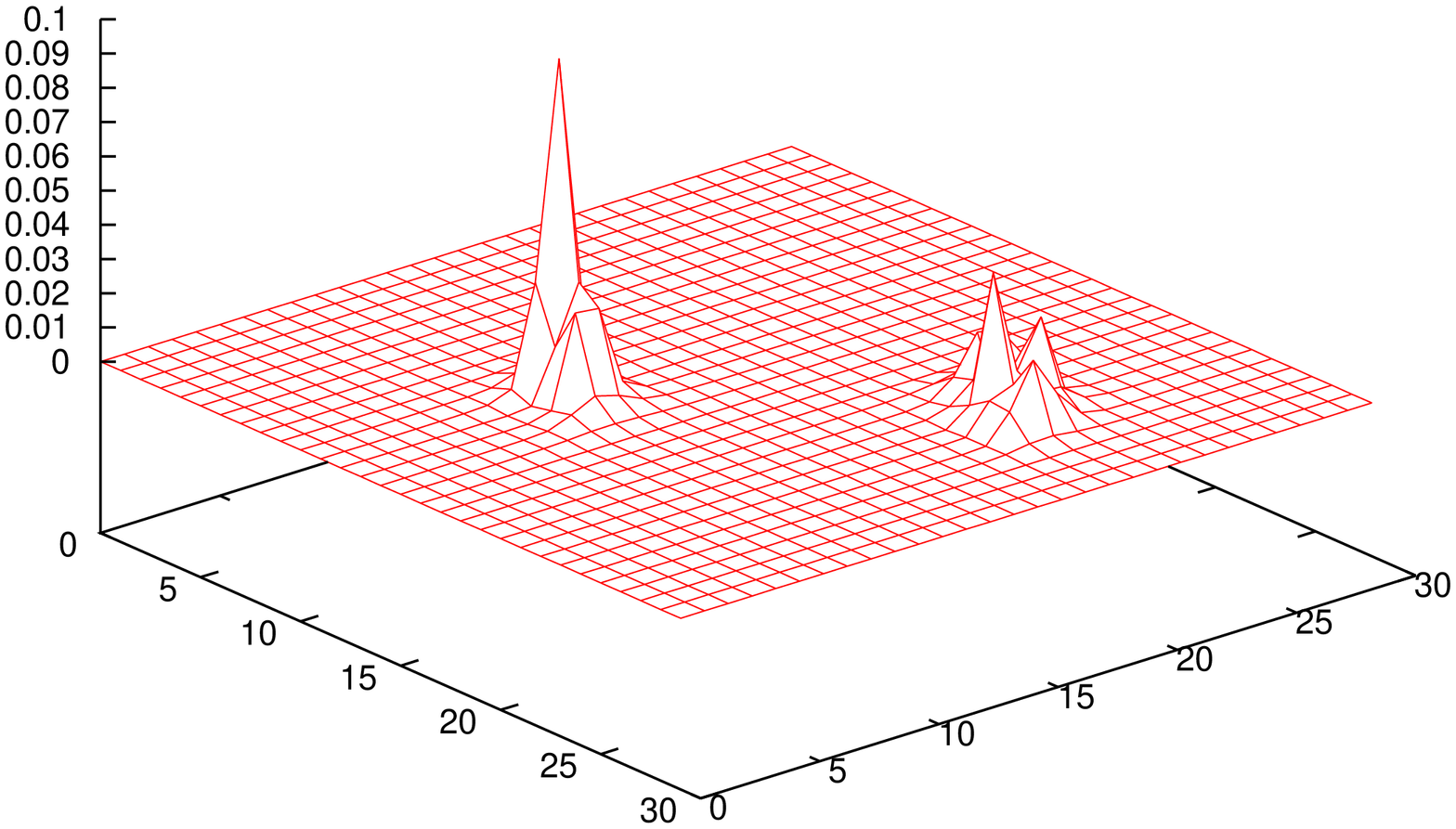}}
\quad 
\subfigure[$E\not=0$, Single valued FT, A sublattice]
{\includegraphics[width=5.0cm,angle=-90]{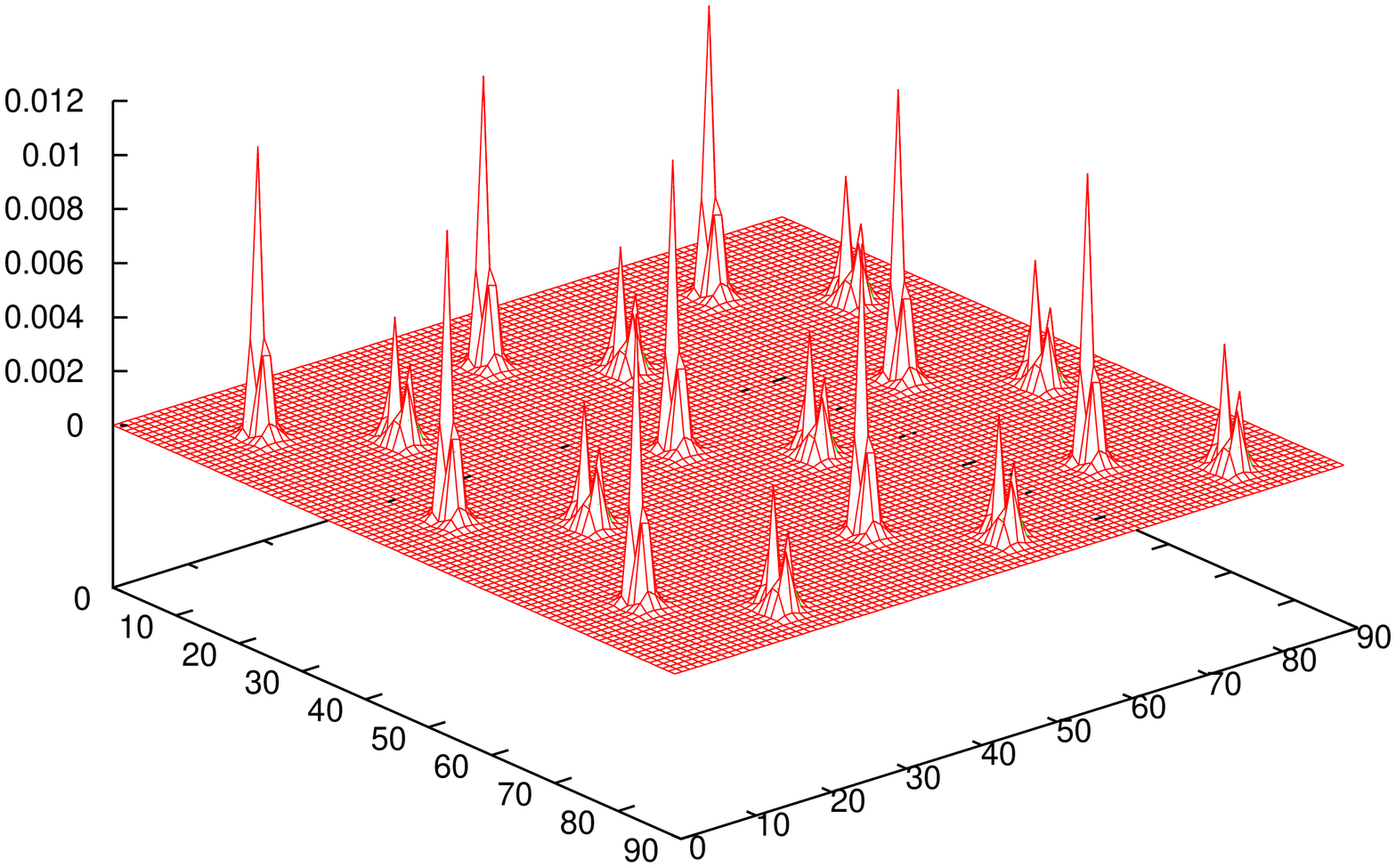}}}
\caption{Eigenvector profiles $\vert\chi\vert^2$ plotted in momentum space.
Results taken on a $30\times30$ lattice with primitive boundary conditions.}
\label{fig:profile}
}  
In order to proceed, recall the discussion of Section~\ref{sec:formulation}, and
in particular that the low energy modes are located in the neighbourhood of the
Dirac points, ie. away from the origin of momentum space. 
It is therefore helpful
to perform analysis in momentum space by Fourier transforming the eigenvectors
$\chi_i(x)$. It turns out that
calculating the discrete Fourier transform on a honeycomb lattice is rather
tricky \cite{alim}, essentially because the range of $k$-values required for a 
unique invertible Fourier transform to exist is larger than is the case for a
square lattice. As shown in the Appendix, we have the choice of defining 
a single-valued transform ranging over either $6L_XL_Y$ (perpendicular)
or $9L_0L_1$ (primitive) modes, or a multi-valued transform ranging over 
$2L_XL_Y$ (perpendicular) or $L_0L_1$ (primitive).
This is exemplified in
Fig.~\ref{fig:profile}, where the single-valued transform shown in
Fig.~\ref{fig:profile}d to a very good approximation consists of nine
copies of the multi-valued transform defined over a smaller range shown in
Fig.~\ref{fig:profile}c. In fact, while we have carried out all subsequent
analysis using both variants of the Fourier transform, the results in all cases 
are found to be identical, as exemplified by Fig.~\ref{maxtopcharge} below.

In Fig.~\ref{fig:profile} we have chosen to transform the eigenvector densities
on A and B sublattices separately to expose an important distinction between
zero and non-zero modes. In each case the modes are tightly localised around two
complementary locations in $k$-space, which we identify with the
$\pm$ Dirac points discussed previously. However, while the non-zero mode of
Figs.~\ref{fig:profile}c,d 
clearly has support at both Dirac points, for the zero
mode the eigenvector on the A sublattice is supported only near the + point 
(Fig.~\ref{fig:profile}a) whereas on the B sublattice it is supported only near
the - point (Fig.~\ref{fig:profile}b).

This behaviour can be understood via the following heuristic argument. Consider
solving the continuum problem on an infinite volume in a more symmetric gauge:
$A_x=-{\omega\over2}y$; $A_y={\omega\over2}x$. Solutions can readily
be found of the form
\be
\psi_{\ell\pm}\propto(x\mp iy)^\ell
\exp\left({-{{\vert\omega\vert}\over4}(x^2+y^2)}\right)\phi_\pm
\label{eq:circ}
\ee
with corresponding eigenvalues
\be
E^2_{\ell\pm}=(\ell+1)[\vert\omega\vert\mp\omega],\;\;\;\ell=0,1,2,\ldots
\ee
The spectrum (\ref{eq:spec_anal},\ref{eq:degen_anal}) is reproduced, with 
both chiralities contributing to non-zero modes, but with only positive
chirality zero-modes present for $\omega>0$ and {\it vice-versa}. However, in
this gauge $\psi_{\ell\pm}$
is also an eigenstate of orbital angular momentum
$\hat L=i(y\partial_x-x\partial_y)$ with eigenvalue $\mp\ell$.
Physically, the $\psi_{\ell\pm}$ describe particles executing circular 
motion (the modes (\ref{eq:circ}) are localised on annular regions 
centred at the origin) with opposite senses for $+$ and $-$ states -
in other words the particle's charge and hence its response to a magnetic field 
is determined by its chirality. Now, in the rest frame both chiralities 
yield orbits of the same shape. However, for our honeycomb fermions the 
states are located at the Dirac points, and hence the previous picture needs to
be Lorentz-boosted. A charged particle moving through a magnetic 
field with non-zero linear momentum has as trajectory 
a 2$d$ projection of a helix;
$+$ and $-$ particle states can no longer
be superimposed and must therefore be described
by different wavefunctions. Hence a state with a well-defined chirality is
necessarily localised around a single point in $k$-space.

We learn from this argument that constructing states with well-defined continuum
quantum numbers may not be straightforward for honeycomb fermions. In
particular, the naive definition of chirality (\ref{eq:naivechiral}) derived for
free fermions appears not to be suitable for the fermion modes in constant
background flux of Fig.~\ref{fig:profile}a,b , since in this case it
receives cancelling contributions from A and B sublattices. Instead, we 
propose the following definition of chirality for interacting honeycomb
fermions:
\be
\bar\psi_\alpha\Sigma_3\psi_\alpha
=-\bar\chi_{B+}\chi_{B+}+\bar\chi_{A+}\chi_{A+}
-\bar\chi_{A-}\chi_{A-}+\bar\chi_{B-}\chi_{B-}.
\label{eq:smartchiral}
\ee
Equation (\ref{eq:smartchiral}) is evaluated in Fourier space, with each mode
counted as $+$ or $-$ depending on which Dirac point it lies closer
to.~\footnote{The
location of the Dirac point 
depends on the boundary conditions and in practice is
determined by the maximum of $\bar\chi\chi$.}
We note in passing that the chirality
operator introduced in the (2+1)-dimensional treatment of \cite{jackiw_pi} also
assigns opposite chiral charges to fields localised around the different Dirac
points.
\FIGURE[ht]{
\includegraphics[width=12.0cm]{chirality_noise.eps}
\caption{Chirality vs. mode number for various random gauge noise $\rho$}
\label{fig:noisy}
}  
In Fig.~\ref{fig:noisy} we plot the expectation values of the chirality 
operator (\ref{eq:smartchiral}) evaluated 
on the lowest 40 eigenmodes of a
30$\times30$ system with $Q=4$. 
We have repeated the calculation, each time
implementing a random gauge transformation of the form $e^{i\theta({\bf
r})\rho}$ at each site,
where $\theta$ is uniformly distributed around the circle and $\rho$ is
a parameter. 
This transformation of course leaves the spectrum unchanged. 
For the smooth untransformed background the chirality
$\langle\Sigma_3\rangle=+1$ to a good approximation for the $2\vert
Q\vert$ topological zero modes. For non-zero modes
the chirality has a smaller magnitude and a fluctuating
sign; moreover
its sum over all degenerate non-zero
modes is exactly zero.
Both of these are of course minimum requirements
for a
realistic chirality operator.
However, Fig.~\ref{fig:noisy} also confirms that the operator 
(\ref{eq:smartchiral}) is not gauge
invariant, which is not surprising since it is formulated in momentum space.
As the amplitude of the short-wavelength noise injected into the gauge
background grows with $\rho$, the
magnitude of $\langle\Sigma_3\rangle$ falls steadily, until eventually the zero
and non-zero modes become indistinguishable. 
\FIGURE[ht]{
\includegraphics[width=12.0cm]{chirality.eps}
\caption{$Q^{\rm index}$ versus $Q^{\rm flux}$ for different lattices}
\label{maxtopcharge}
}  
Using the definition (\ref{eq:smartchiral}) on backgrounds with $\rho=0$
we present results for the topological charge $Q^{\rm
index}$, as
evaluated via the index theorem (\ref{eq:AS}) on the zero modes, 
versus the charge
$Q^{\rm flux}$, obtained by integrating the background
flux (\ref{eq:quant}), for a range of lattice sizes using both boundary
conditions in Fig.~\ref{maxtopcharge}. For $Q$ not too large the curves fall on
a straight line of unit slope independent of lattice volume, confirming the
validity of the definition (\ref{eq:smartchiral}) and verifying the index
theorem. However beyond some value of $Q$, which depends on $L$,
the curves reach a maximum and then
fall with increasing flux density. Different behaviour is observed for the 
two kinds of boundary condition but reassuringly, as mentioned above,
 the results are
insensitive to which definition of the Fourier transform is used.

In fact, the maximal topological charge achievable on a given lattice depends
linearly on the length of its perimeter, rather than the area as naively one
would expect.  For instance in the case of an $L_X\times L_Y$ perpendicular
compactification, provided that that a different scale factor for the X and Y
directions is chosen, 
then the maximum achievable $Q^{\rm index}$ is linearly
proportional to both $L_X$ and $L_Y$ independently.  In particular in
Fig.~\ref{perimeterlaw} we plot $Q^{\rm index}_{max}$ as a function of
``perimeter'' $P=3L_X+{\surd3\over4}L_Y$
for all possible combinations of $L_X$ and $L_Y$ in the range
20-30 independently, plus for other larger lattices with $L_X=L_Y$.
\FIGURE[ht]{
\includegraphics[width=9.0cm,angle=-90]{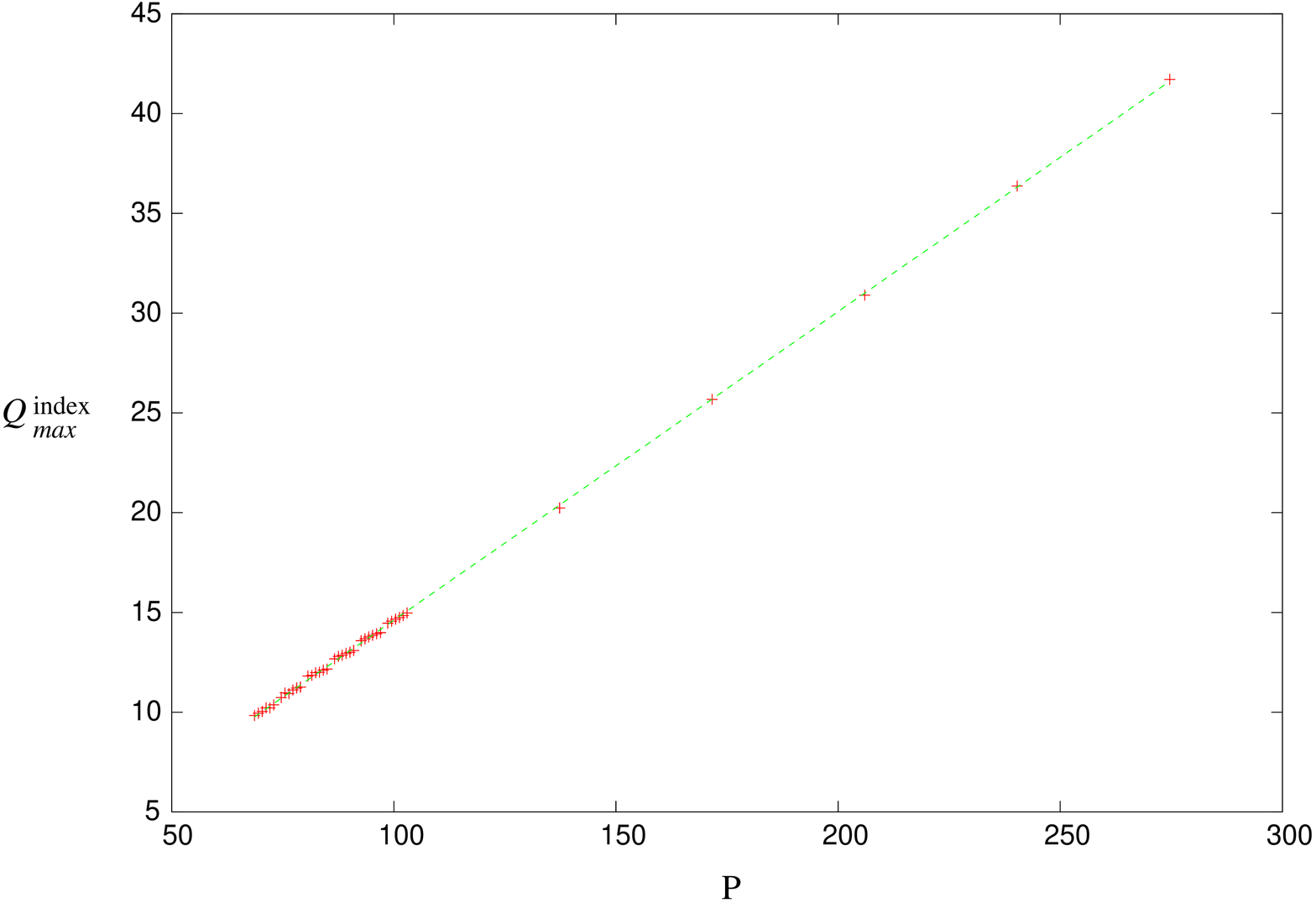}
\caption{$Q^{\rm index}_{max}$ vs. perimeter $P$ in
the perpendicular case} 
\label{perimeterlaw} 
}  
A possible interpretation of this phenomenon is that for 
a $2d$ U(1) gauge theory with constant magnetic flux and fixed $Q$, it is
always possible to perform a gauge transformation that moves all information
about the topological charge in the system 
to the border of the lattice, where it will be
encoded by transition functions $\Omega_x$, $\Omega_y$ such as
(\ref{eq:transition}).  In this case the topological charge can be identified
with the number of windings of a scalar field around the border, consistent with
the quantisation condition (\ref{eq:Qquant}).  It follows that the maximal
resolution obtainable is given by the number of points on the perimeter, i.e.
the ratio of the length of the perimeter to the step size, since we are not able
to probe a field winding a greater number of times than the number of points
defining the discrete Fourier transform.

\section{Concluding Remarks}
\label{sec:conclusions}
In this paper we have demonstrated that the spectrum of a simple fermion model
formulated on a $2d$ honeycomb lattice does indeed reproduce one of the most
important non-perturbative features of relativistic chiral fermions interacting
with a gauge field, namely the index theorem (\ref{eq:AS1}) relating the number
of zero-eigenvalue chiral modes to the background topological charge. We have
done this first by calculating the gauge-invariant spectrum in a particularly
smooth gauge background and showing that it coincides with analytic expectations
yielding zero modes with the correct degeneracy, and next by identifying a
suitable chirality operator $\Sigma_3$ enabling the index to be calculated. The
operator (\ref{eq:smartchiral}) distinguishes between fermion fields located on
differing A and B sublattices in real space, and between fields localised at
differing $\pm$ Dirac points in momentum space. As such it is necessarily 
not gauge-invariant, and hence has limited practical value except in the
artificially-constructed smooth backgrounds used here. Since the spectrum and by
extension $\mbox{det}D$ are gauge-invariant, however, this need not deal a fatal
blow to any simulation programme based on honeycomb fermions.

Since we have constructed the operator (\ref{eq:smartchiral}) to work on a
smooth gauge background, it is legitimate to ask how universal it is, ie. how
would it respond in a non-uniform distribution of topological charge, such as
that found in the vicinity of a vortex line? In the deep continuum limit all
gauge backgrounds can be reduced at least locally to a smooth background by a
suitable gauge transformation, where we know (\ref{eq:smartchiral}) is
appropriate, so it is difficult to see how any alternative definition could be
preferred. Nonetheless, it seems likely that $\langle\Sigma_3\rangle$ as defined
by (\ref{eq:smartchiral}) will be extremely susceptible to lattice artifacts.

We have also studied the maximum $Q^{\rm index}_{max}$ observable in a lattice
simulation, and shown that this scales with the lattice perimeter, arguing that
the limit is related to the maximum resolution of the discrete Fourier transform
along a lattice direction. To our knowledge this is a new observation; it would
be interesting to repeat this analysis for staggered lattice fermions on a
square lattice.

We close by contrasting what we have found to what is known for staggered
lattice fermions. Superficially the two approaches are very similar; one starts
with a single-component Grassmann field on each site and then reallocates the 
degrees of freedom into a new basis to recover fields with spin and flavor
quantum numbers appropriate to continuum fermions. The method used by Smit and
Vink~\cite{smit_vink} partitions the original field in momentum space, according
to a formalism originally developed in \cite{vandenDoel:1983mf}. In that case,
however, the $2^d$ ``Dirac points'' include the origin, and in contrast to the
discussion leading to eqn.  (\ref{eq:contaction}), each continuum flavor is
evenly distributed across all such points. Bilinears with a specified
spin/flavor structure constructed from free fermion fields continue to be
applicable in the presence of gauge fields. Moreover a distinct repartition of
the fields, based on their relative location in local ``hypercubes'' each
containing $2^d$ lattice points, is also possible~\cite{KlubergStern:1983dg},
enabling the construction of local gauge-invariant bilinears by suitable
insertion of products of link variables. For instance
the chirality operator for $2d$ staggered
fermions couples $\chi$ and $\bar\chi$ at
opposite corners of an elementary square~\cite{smit_vink}. 
The two formalisms coincide in the
long-wavelength limit~\cite{Daniel:1986zm}. For honeycomb fermions, the
continuum flavors are localised at different Dirac points. We have argued in
Sec.~\ref{sec:results} that this implies that bilinears appropriate for free
fermions need not continue to be correct once gauge interactions are introduced;
it is also the case that construction of bilinears such as
(\ref{eq:smartchiral}) requires a simultaneous reshuffling both in position and
in momentum space, making gauge invariance impossible to achieve. 

\section*{Acknowledgements}
The work of DC is supported by the European Commission under project number
39494. We have enjoyed useful discussions with Roberto Auzzi and Agostino
Patella.

\appendix
\section{Fourier transform on a honeycomb lattice}
\subsection{Single-valued Fourier transform}
The main result we want to obtain in this section is to write down a unique 
Fourier transform function $\tilde{f}(u,v)$ with a well-defined inverse.
Given the peculiar structure of the hexagonal lattice, it will be shown that 
the range of values of $u$ and $v$ of $\tilde{f}(u,v)$ is larger than for
the square case, and depends on the compactification.

To define a Fourier transform we need to expose which elementary 
translations leave the lattice unchanged; naturally the different types 
of compactification implement these translations differently.
If we want a natural coordinate to enumerate the sites of the lattice we 
need to define a different support function for each compactification type.

We can then define a support function that defines the location of all the 
sites of our lattice $\Pi(x,y)$, by dividing the sites into 
classes, where by a class we mean
the subgroup of sites related by a integer combination of 
elementary translations.
It's straightforward to see that for the perpendicular lattice we need 
four classes to reproduce the entire lattice: 
\be
\Pi_\perp(x,y)
&=&\sum_{n=0,L_X-1}\sum_{m=0,L_Y/2-1}\delta(x-{3\over 2}-3n, y-\sqrt{3}m)
+\delta(x-3n, y-{\sqrt{3}\over 2}-\sqrt{3}m)+ \nonumber \\
&&\delta(x-{1\over 2}-3n, y-\sqrt{3}m) +
\delta(x-2-3n, y-{\sqrt{3}\over 2}-\sqrt{3}m), 
\ee
while for the primitive lattice we need only two classes:
\be
\Pi_p(x,y)&=&\sum_{n=0,L_0-1}\sum_{n=0,L_1-1}\delta(x-\sqrt{3}n, y-\sqrt{3}m)+ \nonumber \\
&&\delta(x-\sqrt{3}n-{\sqrt{3}\over 2}, y-\sqrt{3}n-{1\over\sqrt{3}}).
\ee

The Fourier transform of any function $ f(x,y)$ can then be defined as
\be
\tilde{f}(u,v) =
 {1\over V}
\int \mbox{d}x \mbox{d}y~ \Pi(x,y) ~f(x,y)~ e^{i \vec{p} \cdot \vec{r}},
\ee
where $\vec{p}$ is again different for the two compactifications
\be
\mbox{Perpendicular} &\to& \vec{p}\equiv(\frac{2 \pi}{3 L_X} u~ \hat{X},
\frac{4 \pi}{\sqrt{3}L_Y} v~ \hat{Y})\nonumber \\
\mbox{Primitive} &\to& \vec{p}\equiv(\frac{2 \pi}{\sqrt{3}L_0} u~\hat{0},\frac{2 \pi}{\sqrt{3}L_1} v~\hat{1})
\ee
This leads to the two different definitions of Fourier transform:
\begin{itemize}
\item[]Perpendicular:
\be
\tilde f_\perp(u,v) &=&\frac{1}{\sqrt{6 L_X L_Y}}\sum_{n=0}^{L_X-1}
\sum_{m=0}^{L_Y/2 -1}e^{-2\pi i\Big({n u\over L_X}+{2 m v\over L_Y}\Big)}
\Big[f_\perp({3\over 2}+3n,\sqrt{3}m) e^{-i{\pi u\over L_X}} + \nonumber \\
&& f_\perp(3n, {\sqrt{3}\over 2}+\sqrt{3}m)e^{-i{2\pi v\over L_Y}}+
f_\perp({1\over 2}+3n,-\sqrt{3}m) e^{-i{\pi u\over 3 L_X}}\nonumber \\
&&+f_\perp(2+3n,{\sqrt{3}\over 2}+\sqrt{3}m) e^{-i{4\pi u\over 3 L_X}-
i{2\pi v\over L_Y}}\Big].
\ee
\item[]Primitive:
\be
\tilde f_p(u,v) &=&\frac{1}{3\sqrt{L_0 L_1}}\sum_{n=0}^{L_0-1}\sum_{m=0}^{L_1 -1}e^{-2\pi i\Big({n u\over L_0}+{m v\over L_1}\Big)}
\Big[ f_p(\sqrt{3}n,\sqrt{3}m) + \nonumber \\
&&f_p(\sqrt{3}n+\frac{2}{\sqrt{3}}, \sqrt{3}m +\frac{1}{\sqrt{3}})e^{-i 2\pi\left({2 u\over 3 L_0}-{ v\over3 L_1 }\right)} \Big].
\ee
\end{itemize}
Now note that the periodicity of the transformed functions so defined
differ from the original:
\be
\tilde f_\perp(0,0)&=&\tilde f_\perp(6 L_X, L_Y) \nonumber \\
\tilde f_p(0,0)&=&\tilde f_p(3 L_0,3 L_1). 
\ee
To define the inverse transform we cannot define an unique formula
for all classes. Rather,
the inverse must be calculated on each class separately.
\begin{itemize}
\item[]Perpendicular:
\be
f_\perp({3\over 2}+3n,\sqrt{3}m) &=& \frac{1}{\sqrt{6 L_X L_Y}} 
\sum_{u=0}^{6L_X-1}
\sum_{v=0}^{L_Y -1}
e^{2\pi i\Big({n u\over L_X}+{2 m v\over L_Y}\Big)}\tilde f_\perp(u,v) 
e^{i{\pi u\over L_X}} \nonumber \\
f_\perp(3n, {\sqrt{3}\over 2}+\sqrt{3}m) &=&\frac{1}{\sqrt{6 L_X L_Y}} 
\sum_{u=0}^{6L_X-1}\sum_{v=0}^{L_Y -1}
e^{2\pi i\Big({n u\over L_X}+{2 m v\over L_Y}\Big)}\tilde f_\perp(u,v) 
e^{i{2\pi v\over L_Y}}\nonumber  \\
f_\perp({1\over 2}+3n,-\sqrt{3}m)  &=&\frac{1}{\sqrt{6 L_X L_Y}} 
\sum_{u=0}^{6L_X-1}\sum_{v=0}^{L_Y -1}
e^{2\pi i\Big({n u\over L_X}+{2 m v\over L_Y}\Big)}\tilde f_\perp(u,v) 
e^{i{\pi u\over 3L_X}}\nonumber  \\
f_\perp(2+3n,{\sqrt{3}\over 2}+\sqrt{3}m) &=&\frac{1}{\sqrt{6 L_X L_Y}} 
\sum_{u=0}^{6L_X-1}\sum_{v=0}^{L_Y -1}
e^{2\pi i\Big({n u\over L_X}+{2 m v\over L_Y}\Big)}\tilde f_\perp(u,v) 
e^{i{4\pi u\over 3L_X}+i{2\pi v\over L_Y}}\nonumber  \\
\ee
\item[]Primitive:
\be
f_p(\sqrt{3}n,\sqrt{3}m) &=& \frac{1}{3\sqrt{L_0 L_1}} \sum_{u=0}^{3L_0-1}
\sum_{v=0}^{3L_1 -1}
e^{2\pi i\Big({n u\over L_0}+{m v\over L_1}\Big)}\tilde f_p(u,v)\nonumber \\
f_p(\sqrt{3} n+ {2 \over \sqrt{3}},\sqrt{3}m+ {1 \over \sqrt{3}}) &=&\nonumber\\
\frac{1}{3\sqrt{L_0 L_1}} 
\sum_{u=0}^{3L_0-1}\sum_{v=0}^{3L_1-1} &&\hspace{-3mm}
e^{2\pi i\Big({n u\over L_0}+{m v\over L_1}\Big)}\tilde f_p(u,v) 
e^{2\pi i\Big({2 u\over 3 L_0}+{v\over 3 L_1}\Big)}
\ee
\end{itemize}
\subsection{Multi-valued Fourier transform}
If we relax the requirement of having a single-valued function of momentum, 
but still wish to expose the different structure of the  A and B sites 
in the Fourier transform, we obtain two further definitions.
In this case the the range of values of $u$ and $v$ depends on the 
compactification, but is always smaller than the previous case.
The procedure to obtain these function is totally equivalent to the previous 
case, so here we only show the results.
\begin{itemize}
\item[]Perpendicular:
\be
\tilde f^{(A)}_\perp(u,v) &=&\frac{1}{\sqrt{2 L_X L_Y}}\sum_{n=0}^{L_X-1}
\sum_{m=0}^{L_Y/2 -1}e^{-2\pi i\Big({n u\over L_X}+{2 m v\over L_Y}\Big)}
\Big[f_\perp({3\over 2}+3n,\sqrt{3}m) e^{-i{\pi u\over L_X}} + \nonumber \\
&& f_\perp(3n, {\sqrt{3}\over 2}+\sqrt{3}m)e^{-i{2\pi v\over L_Y}}\Big]\nonumber \\
\tilde f^{(B)}_\perp(u,v) &=&\frac{1}{\sqrt{2 L_X L_Y}}\sum_{n=0}^{L_X-1}
\sum_{m=0}^{L_Y/2 -1}e^{-2\pi i\Big({n u\over L_X}+{2 m v\over L_Y}\Big)}
\Big[f_\perp({1\over 2}+3n,-\sqrt{3}m) e^{-i{\pi u\over 3 L_X}}\nonumber \\
&&+f_\perp(2+3n,{\sqrt{3}\over 2}+\sqrt{3}m) e^{-i{4\pi u\over 3 L_X}-
i{2\pi v\over L_Y}}\Big].
\ee
\item[]Perpendicular Inverse:
\be
f_\perp({3\over 2}+3n,\sqrt{3}m) &=& \frac{1}{\sqrt{2 L_X L_Y}} 
\sum_{u=0}^{2L_X-1}\sum_{v=0}^{L_Y -1}
e^{2\pi i\Big({n u\over L_X}+{2 m v\over L_Y}\Big)}\tilde f^{(A)}_\perp(u,v) 
e^{i{\pi u\over L_X}}\nonumber \\
f_\perp(3n, {\sqrt{3}\over 2}+\sqrt{3}m) &=&\frac{1}{\sqrt{2 L_X L_Y}} 
\sum_{u=0}^{2L_X-1}\sum_{v=0}^{L_Y -1}
e^{2\pi i\Big({n u\over L_X}+{2 m v\over L_Y}\Big)}\tilde f^{(A)}_\perp(u,v) 
e^{i{2\pi v\over L_Y}}\nonumber \\
f_\perp({1\over 2}+3n,-\sqrt{3}m)  &=&\frac{1}{\sqrt{2 L_X L_Y}} 
\sum_{u=0}^{2L_X-1}\sum_{v=0}^{L_Y -1}
e^{2\pi i\Big({n u\over L_X}+{2 m v\over L_Y}\Big)}\tilde f^{(B)}_\perp(u,v) 
e^{i{\pi u\over 3L_X}}\nonumber\\
f_\perp(2+3n,{\sqrt{3}\over 2}+\sqrt{3}m) &=&\frac{1}{\sqrt{2 L_X L_Y}} 
\sum_{u=0}^{2L_X-1}\sum_{v=0}^{L_Y -1}
e^{2\pi i\Big({n u\over L_X}+{2 m v\over L_Y}\Big)}\tilde f^{(B)}_\perp(u,v) 
e^{i{4\pi u\over 3L_X}+i{2\pi v\over L_Y}}\nonumber\\ 
\ee
\item[]Primitive:
\be
\tilde f^{(A)}_p(u,v) &=&\frac{1}{\sqrt{L_0 L_1}}\sum_{n=0}^{L_0-1}
\sum_{m=0}^{L_1 -1}e^{-2\pi i\Big({n u\over L_0}+{m v\over L_1}\Big)}
f_p(\sqrt{3}n,\sqrt{3}m)\nonumber \\
\tilde f^{(B)}_p(u,v) &=&\nonumber \\
\frac{1}{\sqrt{L_0 L_1}}&&\hspace{-4mm}\sum_{n=0}^{L_0-1}
\sum_{m=0}^{L_1 -1}e^{-2\pi i\Big({n u\over L_0}+{m v\over L_1}\Big)}
f_p(\sqrt{3}n+\frac{2}{\sqrt{3}}, \sqrt{3}m +\frac{1}{\sqrt{3}})
e^{-i 2\pi\left({2 u\over 3 L_0}-{ v\over3 L_1 }\right)} \nonumber\\ 
\ee

\item[]Primitive Inverse:
\be
f_p(\sqrt{3}n,\sqrt{3}m) &=&\frac{1}{\sqrt{L_0 L_1}} \sum_{u=0}^{L_0-1}
\sum_{v=0}^{L_1 -1}
e^{2\pi i\Big({n u\over L_0}+{m v\over L_1}\Big)}\tilde f^{(A)}_p(u,v)\nonumber \\
f_p(\sqrt{3} n+ {2 \over \sqrt{3}},\sqrt{3}m+ {1 \over \sqrt{3}}) &=&\nonumber \\
\frac{1}{\sqrt{L_0 L_1}} &&\hspace{-4mm}
\sum_{u=0}^{L_0-1}\sum_{v=0}^{L_1-1}
e^{2\pi i\Big({n u\over L_0}+{m v\over L_1}\Big)}\tilde f^{(B)}_p(u,v) 
e^{2\pi i\Big({2 u\over 3 L_0}+{v\over 3 L_1}\Big)}\nonumber\\
\ee
\end{itemize}
\newpage



\begin{thebibliography}{99}
\bibitem{KarstenWilczek}
  L.H.~Karsten,
  Phys.\ Lett.\  B {\bf 104}, 315 (1981);\\
  F.~Wilczek,
  Phys.\ Rev.\ Lett.\  {\bf 59}, 2397 (1987).

\bibitem{creutz}
  M.~Creutz,
  JHEP {\bf 0804}, 017 (2008).

\bibitem{borici}  A.~Bori\c ci,
  Phys.\ Rev.\  D {\bf 78}, 074504 (2008).

\bibitem{bedaque}
  P.F.~Bedaque, M.I.~Buchoff, B.C.~Tiburzi and A.~Walker-Loud,
  Phys.\ Lett.\  B {\bf 662}, 449 (2008).

\bibitem{Cichy}
  K.~Cichy, J.~Gonzalez Lopez, K.~Jansen, A.~Kujawa and A.~Shindler,
  Nucl.\ Phys.\  B {\bf 800} (2008) 94.

\bibitem{NN}
  H.B.~Nielsen and M.~Ninomiya,
  Nucl.\ Phys.\  B {\bf 185}, 20 (1981)
  [Erratum-ibid.\  B {\bf 195}, 541 (1982)];
  Nucl.\ Phys.\  B {\bf 193}, 173 (1981).

\bibitem{creutz2}
  M.~Creutz,
  arXiv:0808.0014 [hep-lat].

\bibitem{smit_vink} 
J.~Smit and J.~C.~Vink,
  Nucl.\ Phys.\  B {\bf 286}, 485 (1987).

\bibitem{bedaque2}
  P.F.~Bedaque, M.I.~Buchoff, B.C.~Tiburzi and A.~Walker-Loud,
 Phys.\ Rev.\  D {\bf 78}, 017502 (2008).

\bibitem{Semenoff:1984dq}
  G.W.~Semenoff,
  Phys.\ Rev.\ Lett.\  {\bf 53} (1984) 2449.

\bibitem{Jackiw:1984ji}
  R.~Jackiw,
  Phys.\ Rev.\  D {\bf 29} (1984) 2375
  [Erratum-ibid.\  D {\bf 33} (1986) 2500].

\bibitem{QHE} 
V.P.~Gusynin and S.G.~Sharapov,
  Phys.\ Rev.\ Lett.\  {\bf 95} (2005) 146801;\\
Y.~Hatsugai, T.~Fukui and H.~Suzuki, Phys.\ Rev. \ B {\bf 74},
205414 (2006). 

\bibitem{Stone} J.K. Pachos and M. Stone, 
Int.\ J. Mod. Phys. {\bf B21} (2007) 5113.

\bibitem{jackiw_pi} R.~Jackiw and S.~Y.~Pi,
 Phys.\ Rev.\ Lett. {\bf 98}, 266402 (2008).

\bibitem{saad} Y. Saad, {\it Numerical Methods For Large Eigenvalue Problems},
Manchester University Press, 1992.

\bibitem{luscher}
  L.~Del Debbio, L.~Giusti, M.~L\"uscher, R.~Petronzio and N.~Tantalo,
  JHEP {\bf 0602}, 011 (2006).

\bibitem{alim} 
  U.R.\ Alim and T.~M\"oller,
  {\it A discrete Fourier transform 
for the hexagonal and body-centered lattices}, 
SFU Computing Science Technical Report 2008-14;\\
  A.\ Vince and X.\ Zheng, J.\ Math.\ Imaging\ Vis. {\bf 28}, 125 (2007).

\bibitem{vandenDoel:1983mf}
  C.~van den Doel and J.~Smit,
  Nucl.\ Phys.\  B {\bf 228}, 122 (1983).

\bibitem{KlubergStern:1983dg}
  H.~Kluberg-Stern, A.~Morel, O.~Napoly and B.~Petersson,
  Nucl.\ Phys.\  B {\bf 220}, 447 (1983).

\bibitem{Daniel:1986zm}
  D.~Daniel and T.D.~Kieu,
  Phys.\ Lett.\  B {\bf 175}, 73 (1986).

\end{thebibliography}
\end{document}